\documentclass[
reprint,
amsmath,amssymb,
aps,
prb,
]{revtex4-2}

\usepackage{hyperref}
\hypersetup{colorlinks=true,
            linkcolor=black,
            citecolor=black,
            urlcolor=blue}
\usepackage{amssymb}
\usepackage{mathbbol}
\usepackage{courier}

\newcommand{\vb}[1]{\textcolor{black}{#1}}
\newcommand{\new}[1]{\textcolor{black}{#1}}
\newcommand{\krst}[1]{\textcolor{black}{#1}}
\newcommand{\ff}[1]{\textcolor{black}{#1}}

\usepackage{graphicx}% Include figure files
\usepackage{dcolumn}% Align table columns on decimal point
\usepackage{bm}% bold math
\begin{document}

\title{Spin-1 Haldane chains of superconductor-semiconductor hybrids}

\author{Virgil V. Baran$^{1,2,3}$}
\email[]{virgil.v.baran@unibuc.ro}
\author{Jens Paaske$^{3}$}
\affiliation{$^{1}$Faculty of Physics, University of Bucharest, 405 Atomi\c stilor, RO-077125, Bucharest-M\u agurele, Romania}
\affiliation{$^{2}$``Horia Hulubei" National Institute of Physics and Nuclear Engineering, 30 Reactorului, RO-077125, Bucharest-M\u agurele, Romania}
\affiliation{$^{3}$Center for Quantum Devices, Niels Bohr Institute, University of Copenhagen, 2100 Copenhagen, Denmark}

\begin{abstract}
We theoretically explore the possibility of realizing the symmetry-protected topological Haldane phase of spin-1 chains in a tunable hybrid platform of superconducting islands (SIs) and quantum dots (QDs). Inspired by recent findings suggesting that an appropriately tuned QD-SI-QD block may behave as a robust spin-1 unit, we study the behavior of many such units tunnel-coupled into linear chains. Our efficient and fully microscopic modeling of long chains with several tens of units is enabled by the use of the surrogate model solver [Phys. Rev. B 108, L220506 (2023); arXiv:2402.18357]. Our numerical findings indicate that the QD-SI-QD chains exhibit emblematic features of the Haldane phase, such as fractional spin-1/2 edge states and non-vanishing string order parameters, and that these persist over a sizeable region of parameter space. 
%Increasing the coupling between neighboring units gradually degrades their individual spin-1 character and leads to a trivial dimerized phase. 
%As tunnel-coupling multiple QDs to the same SI would lead to robust higher-spin units, in light of our results the super-semi hybrid platform appears promising for realizing the rich phase diagrams of more general Heisenberg spin models.
\vspace*{15mm}

\end{abstract}

\maketitle

\section{Introduction}

Fueled by open problems in both fundamental and applied physics, the field of superconductor-semiconductor hybrids has witnessed sustained advances over the past few decades. Crucial to the understanding of these super-semi systems is the hybridization between their various constituents, which often leads to the presence of tunable subgap states~\cite{Bauer2007Nov, Meng2009Jun, Oguri2013, Yu1965, Shiba1968, Rusinov1969, Kirsanskas2015}.
%They range from Yu-Shiba-Rusinov states~\cite{Yu1965, Shiba1968, Rusinov1969}, induced by Coulomb blockaded QDs carrying a local magnetic moment, to a localized quasiparticle excitation above an induced gap on proximitized QDs with smaller charging energy~\cite{Bauer2007Nov, Meng2009Jun, Oguri2013, Kirsanskas2015}. 
Their properties directly influence the design of superconducting qubits and other complex gateable devices for quantum technologies~\cite{Mishra2021Dec, Pavesic2022Feb, Pita-Vidal2023May}. 
%In good analogy with the formation of a simple hydrogen molecule, t

In particular, the ongoing efforts of realizing poor man's Majorana bound states in short Kitaev chains~\cite{Leijnse2012Oct,Sau2012Jul,Tsintzis2022Nov, Liu2022Dec,Dvir2023Feb, Zatelli2023Nov, Tsintzis2024Feb, Bordin2024Feb} rely on the hybridization of two spatially separated QDs with an extended gateable super-semi subgap state.
%, whose signatures have already been detected in experiments \cite{Su2017Sep,Kurtossy2021Oct,Junger2023Jul}. 
A closely related configuration, where the two QDs are coupled through a floating superconducting island (SI), was recently considered for its interesting exchange properties~\cite{Bacsi2023Sep,Baran2024Feb}. Remarkably, it turns out that the QD-SI-QD \vb{(DSD)} exhibits a robust spin-1 ground state when the QDs couple strongly and coherently through (at least two subgap states in) the SI, in the presence of a sizeable SI Coulomb energy~\cite{Baran2024Feb}. Inspired by this finding, we explore here the possibility of realizing the well-known Haldane phase of spin-1 chains in this super-semi hybrid platform.

%Note that recent experiments indicate that the spatial extension of these states is large and can reach even the range of 50nm \cite{Scherubl2020Apr}.

The Haldane phase is a celebrated symmetry-protected topological phase of matter realized in the gapped ground state of the spin-1 antiferromagnetic Heisenberg chain \cite{Haldane1983Feb,Haldane1983Apr,Pollmann2012Feb,Haldane2017Oct,Wen2017Dec}.  The hallmark of such phases is the existence of \new{particular edge modes} which enjoy some degree of topological protection, i.e. they are robust to symmetry-preserving local perturbations. The Haldane phase of a \new{long-enough} spin-1 chain with open boundary conditions features \new{a four-fold degenerate many-body ground state with the edge modes behaving} as two effective spin-1/2 degrees of freedom despite the fact that the model's elementary building blocks were spins-1. This symmetry \emph{fractionalization} phenomenon \cite{Chen2011Jan} is \new{well-illustrated by the analytical valence bond solid solution of the Affleck-Kennedy-Lieb-Tasaki (AKLT) model \cite{Affleck1987Aug}}.

%, provides a vivid example of Anderson's \emph{more is different} paradigm \cite{Anderson1972Aug}.

The existence of edge states in spin-1 chains has been extensively investigated over the past decades, both theoretically \cite{White1993Aug,Jolicoeur2019,Kim2000Dec} and experimentally \cite{Zaliznyak2001Jun,Kenzelmann2002Jun,Kenzelmann2003Feb,Sompet2022Jun,Nag2022Apr,Jelinek2023Jan}. Various platforms for realizing synthetic spin-1 chains have been proposed over the years, e.g. by using gated triple quantum dots \cite{Shim2010Nov}, arrays
of semiconductor QDs in a nanowire \cite{Jaworowski2017Jul,Manalo2021Sep,Manalo2024Feb}, a chain of triangular graphene QDs \cite{Mishra2021Oct,Catarina2022Feb}, in addition to molecular \cite{Williams2020Jan} and organometallic platforms \cite{Zheng2006Nov,Stone2007Aug, Pitcairn2023Jan,Tin2023Sep}. Various applications
of the spin-1/2 edge states as qubits have also been suggested \cite{Shim2010Nov,Jaworowski2017Jul, Jaworowski2019Jan}. \vb{Furthermore, they have direct implications for measurement-based quantum computation \cite{Miyake2010Jul,Bartlett2010Sep}.}

The purpose of this work is to \vb{show} that the Haldane phase may also be realized in a tunable DSD chain under experimentally reasonable assumptions.  The rest of the paper is organized as follows. In Sec.~\ref{sec:method}, we lay down the modeling methodology based on the surrogate model solver \cite{Baran2023Dec,Baran2024Feb}. Furthermore, we introduce the Heisenberg Hamiltonian and its bilinear-biquadratic generalization as ideal spin-1 chains to be used as benchmarks for the DSD results. In Sec.~\ref{sec:results}, we discuss the numerical results (energy spectra, spin-densities, string order parameters) obtained for DSD chains of increasing length: the individual unit ($N=1$), the dimer ($N=2$), intermediate length chains ($N=3,...12$) and finally long chains ($N=21,41$), which are expected to visibly display the main features of the Haldane phase. We draw conclusions and discuss possible generalizations of our work in Sec.~\ref{sec:conclusions}.

\section{Modeling methodology}
\label{sec:method}

\subsection{DSD chains}

The total Hamiltonian for a length-$N$ DSD chain is given by
\begin{equation}\label{H_chain}
    \hat{H}=\sum_{n=1}^N\hat{H}_{\text{DSD,n}}+t_d \sum_{n=1}^{N-1}\sum_{ \sigma=\uparrow \downarrow}( d_{ nR\sigma}^{\dagger} d_{n+1,L,\sigma}+\text {h.c.})~,
\end{equation}
where $d^\dagger_{n\alpha\sigma}$ creates an electron with spin $\sigma$ in the $\alpha=\text{L,R}$ quantum dot of the $n$-th DSD unit. We consider all units to be identical and described by a Hamiltonian of the form 
\begin{equation}\label{H_DSD}
    \hat{H}_{\text{DSD}}=\hat{H}_{\text{QD,L}}+\hat{H}_{\text{SI}}+\hat{H}_{\text{QD,R}}+\hat{H}_{\text{tunn}}~.
\end{equation}

For each quantum dot $\alpha=\text{L,R}$ we use the \krst{constant interaction} Hamiltonian
\begin{equation}\label{H_QD}
\hat{H}_{\text{QD},\alpha}=\frac{U}{2} \left(  d_{\alpha \uparrow}^{\dagger} d_{\alpha \uparrow}+d_{\alpha \downarrow}^{\dagger} d_{\alpha \downarrow}-\nu\right)^2~,
\end{equation}
where $U$ is the electron-electron repulsion strength and $\nu$ is the dot energy level in units of
electron number. Throughout this work, all QDs are assumed to be identical and to effectively host one electron each, i.e. we always take $\nu=1$ and large enough values for $U$. Departing from this particle-hole symmetric point is known to reduce the excitation gap above the spin-1 DSD ground state of interest here~\cite{Bacsi2023Sep}. Furthermore, we neglect any cross capacitances between the different parts of the DSD units, the effects of which can be accounted for by rescaling the various parameters.

For modeling the SI \vb{and its tunnel couplings} we employ the surrogate model solver (SMS) methodology outlined in Refs.~\onlinecite{Baran2023Dec,Baran2024Feb}. \krst{For a vanishing SI charging energy $E_c$, the full quasi-continuum of SI levels is replaced within the SMS approach by a small number of BCS surrogate orbitals that optimally reproduce the SI-QD hybridization function \cite{Baran2023Dec}. The SMS approach may be generalized to a non-vanishing $E_c$ provided that the finite-size effects of the SI can safely be neglected \cite{Baran2024Feb}. In this case, the SMS prescription consists of coupling the BCS surrogate orbitals (used as before to model the hybridization part) with an auxiliary Cooper pair counting site that enables the conservation of the total SI particle number. This leads to an accurate description of the SI's charge fluctuations and thus ensures a proper treatment of its Coulomb interaction term. Concretely, the SI Hamiltonian is given by}
\begin{equation}\label{H_SI}
\begin{aligned}
    \hat{H}_{\text{SI}} =&\sum_{\ell=1}^{\tilde{L}} \sum_{\sigma=\uparrow \downarrow} \tilde\xi_{ \ell} c_{\ell \sigma}^{\dagger} c_{\ell \sigma}- \sum_{\ell=1}^{\tilde L}(\Delta c_{\ell\uparrow}^{\dagger} c_{\ell \downarrow}^{\dagger} e^{-i\hat{\phi}}+\text{h.c.})\\
    &+E_c \left(\sum_{\ell=1}^{\tilde{L}} \sum_{\sigma=\uparrow \downarrow} c_{\ell \sigma}^{\dagger} c_{\ell \sigma}+2\hat{N}_p-n_0\right)^2~.
    \end{aligned}
\end{equation}
\begin{figure}[ht!]
\includegraphics[width=\columnwidth]{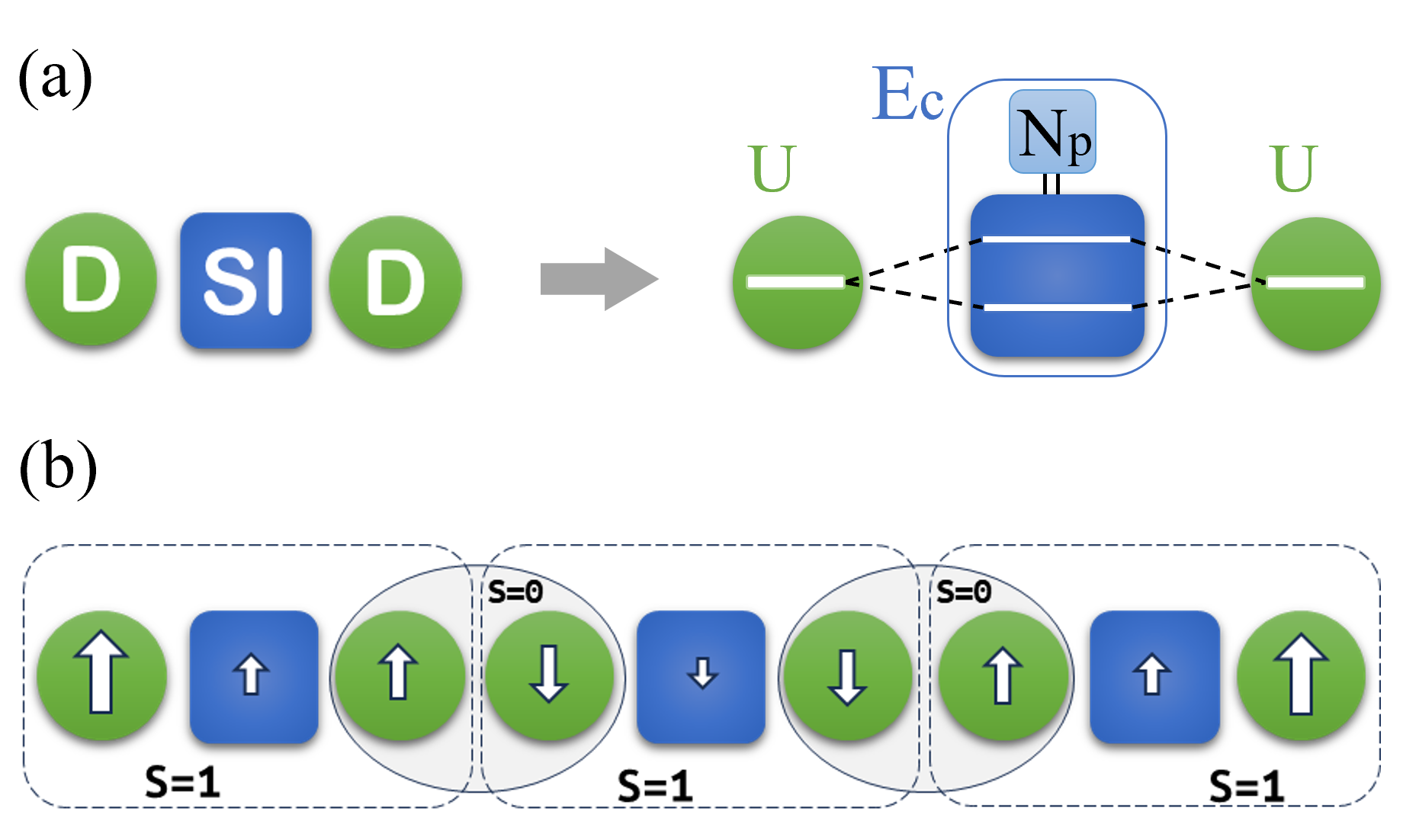}
\caption{(a) Modeling of the DSD unit in the SMS approach with a minimal $\tilde{L}=2$ surrogate for the SI coupled to an auxiliary site counting the number of Cooper pairs $N_p$ in the superconducting condensate. (b) Schematic of the $N=3$ DSD chain in its $S_{\text{tot}}=S_{z,\text{tot}}=1$ ground state. The strength of the $S=0$ valence bond is dictated by the inter-dot tunnel matrix element $t_d$, see Eq.~(\ref{H_chain}).}
\label{fig_1}
\end{figure}
Here $c^\dagger_{i\sigma}$ creates an electron with spin $\sigma$ and energy $\tilde{\xi}_{i}$ in the SI with charging energy $E_c$ and optimal occupation (in units of electron charge) $n_0$. As discussed above, we need to make use of the canonically conjugate number and phase operators $\hat{N}_p$ and $\hat{\phi}$, $[\hat{N}_p,e^{i\hat{\phi}}]=e^{i\hat{\phi}}$. Physically, $\hat{N}_p$ counts the number of Cooper pairs in the superconducting condensate, while $e^{\pm i\hat{\phi}}$ adds/removes one pair from the condensate. The auxiliary Hilbert space is spanned by states $|p\rangle$ with an integer number of pairs \new{$p$}, obeying $\hat{N}_p|p\rangle=p|p\rangle$ and $e^{\pm i\hat{\phi}}|p\rangle=|p\pm 1\rangle$.

Finally, the QD-SI coupling term appearing in Eq.~(\ref{H_DSD}) reads
\begin{equation}
\label{H_tunn}
\hat{H}_{\text{tunn}}=\sum_{\alpha=\text{L,R}}\sum_{\ell=1}^{\tilde L}\sum_{ \sigma=\uparrow \downarrow}\sqrt{\gamma_\ell \Gamma_\alpha}( c_{ \ell \sigma}^{\dagger} d_{\alpha\sigma}+\text {h.c.})~,
\end{equation}
where $\Gamma_\alpha$ denote the QD$_\alpha$-SI tunneling rates. The $\gamma_\ell$ parameters define the surrogate model, together with the energy levels $\tilde{\xi}_\ell$. 

\new{As detailed in Ref.~\onlinecite{Baran2024Feb}, the minimal prescription for reproducing the spin-1 ground state of the DSD unit relies on the $\tilde{L}=2$ surrogate. This is also in agreement with Ref.~\onlinecite{Bacsi2023Sep}, where a finite-bandwidth was found to be essential for this purpose (see also the Appendix for a complementary discussion).} For the \vb{minimal} $\tilde{L}=2$ surrogate employed throughout this work, the numerical values of the above parameters are $\gamma_{1,2}=1.246\Delta$ and $\tilde{\xi}_{1,2}=\pm1.31\Delta$ (obtained by the optimization method detailed in Ref.~\onlinecite{Baran2023Dec} for a half-bandwidth $D=10\Delta$). \new{For these values, the excitation energy of the BCS quasiparticles becomes $E_{\text{qp}}=(\Delta^2+\tilde\xi^2)^{1/2}\simeq 1.65\Delta$.} \vb{More complex $\tilde{L}\geq 3$ surrogates were found to cause only minor quantitative differences for the DSD chains considered below.}

\subsection{Mapping to ideal spin-1 chains}
\label{idealchains}

When the inter-dot coupling $t_d$ is small enough and the picture of robust spin-1 DSD units holds, the low-energy spectrum of $\hat{H}$ \vb{in Eq.~(\ref{H_chain})} may be matched to that of the spin-1 antiferromagnetic Heisenberg chain (AFH) given by
\begin{equation}
    \label{afh}
    \hat{H}_{\text{AFH}}=J_{\text{AF}}\sum_{i=1}^{N-1}\vec{S}_{i}\cdot \vec{S}_{i+1}~,
\end{equation}
with the coupling $J_{\text{AF}}>0$ following the superexchange scaling \cite{Anderson1959Jul} $J_{\text{AF}}\sim t_d^2/U$. The spin-1 character of each DSD unit is gradually lost with increasing $t_d$ 
as the correlations between QDs belonging to neighboring units build up (see also Figs.~\ref{fig_3}a and~\ref{fig_5}b below).

The large-$t_d$ limit \vb{corresponds to a dimerized configuration where} all double-dots are effectively locked into spin-singlet configurations (see Fig.~\ref{fig_1}b), leaving two isolated (dangling) spin-$1/2$ moments at the chain's edges. \vb{Naively}, this behaviour would appear reminiscent of the  AKLT ansatz in which every spin-1
is identified with the triplet subspace of two virtual spins-1/2 \vb{each participating in a singlet bond with its other neighboring spin} \cite{Affleck1987Aug}. The AKLT state is the ground state of the bilinear-biquadratic (BLBQ) Hamiltonian
\begin{equation}
    \label{blbq}
    \hat{H}_{\text{BLBQ}}=J_{\text{AF}}\sum_{i=1}^{N-1}\left[\vec{S}_{i}\cdot \vec{S}_{i+1}+\beta (\vec{S}_{i}\cdot \vec{S}_{i+1})^2\right]~,
\end{equation}
for $\beta_{\text{AKLT}}=1/3$. For this particular value of $\beta$, $\hat{H}_{\text{BLBQ}}$ becomes a sum of projectors onto local spin-2 pairs, which thus favors the formation of spin-singlet valence bonds (in the picture of the spin-1 consisting of two symmetrized virtual spins-1/2). Note that the biquadratic exchange coupling is also the simplest local term that is compatible with all the system's symmetries which may be added to the Heisenberg Hamiltonian of Eq.~(\ref{afh}). The BLBQ model features both gapped excitations and fractional spin-1/2 edge states, with a fourfold degeneracy in the thermodynamic limit, in a range that includes $0\leq \beta \leq 1/3$ belonging to the Haldane phase \cite{Kennedy1990Jul,White1993Aug}.

In the following section, we will argue that the low-energy physics of the DSD chains (\vb{at weak enough $t_d$}) is well captured by the BLBQ Hamiltonian of Eq.~(\ref{blbq}) with $0<\beta<1/3$, and satisfies the necessary requirements of the Haldane phase.

\section{Results}
\label{sec:results}

To obtain the low-lying spectrum of the quasi-1D and locally-interacting systems considered here, our numerical method of choice is the density matrix renormalization group (DMRG) in the matrix-product-state formulation~\cite{White1992Nov, Schollwock2011Jan}, which is straightforward to implement with the ITensor library~\cite{itensor,itensor-r0.3}. Our numerical codes are available online~\cite{github} and may be run on a standard laptop or desktop computer. We employed a maximum bond dimension of 2000 and an energy convergence threshold of $5\cdot 10^{-9}\Delta$. We truncated the auxiliary Hilbert space for each SI to the dimension $d_{\text{aux}}=10$. This relatively small value of $d_{\text{aux}}$ is able to account for all relevant SI charge fluctuations given the relatively large charging energy $E_c=2\Delta$ chosen in our simulations. The longest running times were of the order of a few days for the longest $N=41$ DSD chain discussed below, whose total Hilbert space dimension would amount to $(4^{\tilde{L}+2}\cdot d_{\text{aux}})^N \sim 10^{140}$  for the $\tilde{L}=2$ surrogate.

\subsection{The DSD unit}

\begin{figure}[ht!]
\includegraphics[width=\columnwidth]{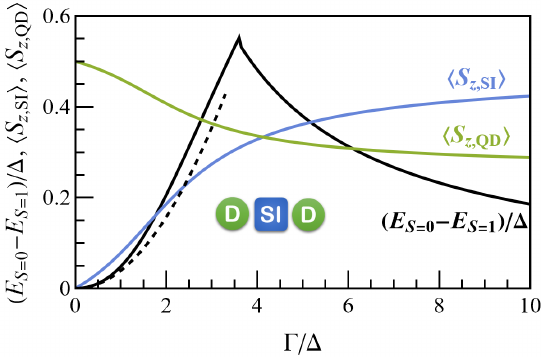}
\caption{DSD spin-singlet-triplet energy gap and average SI and QD spins $\langle S_z\rangle$ in the $S_{\text{tot}}=S_{z,\text{tot}}=1$ ground state. \vb{The dashed line indicates the spin-singlet-triplet gap obtained by perturbation theory (see Appendix).} The other parameters are $U=-2\epsilon_d=6\Delta$, $E_c=2\Delta$. For the SI we employed the $\tilde{L}=2$ surrogate.}
\label{fig_2}
\end{figure}

Let us first shortly revisit the elementary QD-SI-QD (DSD) unit, whose properties have been investigated in some detail in Refs.~\onlinecite{Bacsi2023Sep,Baran2024Feb}. The emergence of the spin-triplet ground state in this setup may be understood in analogy to the double-dot configuration of Ref.~\onlinecite{Probst2016Oct}. In a perturbative approach, in addition to the hybridization between the QDs' spin-singlet and the Cooper pairs (that would naively lead to an overall singlet ground state), one must consider 4$^{th}$ order tunneling processes involving single-particle excitations in the superconducting leads. When it becomes advantageous to perform
the spin-exchange by virtually exciting the superconduc-
tor instead of the QDs, a spin-triplet ground state may
emerge, as detailed in the Appendix and in Ref.~\onlinecite{Probst2016Oct}.

In modeling the DSD unit, we implicitly assumed that the QDs are coupled through the \emph{same} SI-orbitals, much like in the poor-man's Majorana devices where a spatially extended state in the proximitized semiconductor allows for crossed Andreev reflections~\cite{Tsintzis2022Nov,Liu2022Dec,Dvir2023Feb}.  Also, it is important to note that the spin-triplet character of the QD-SI-QD system is rather sensitive to the asymmetry in the magnitude of the couplings between the QD and superconducting levels (see discussion around Fig.~8 of Ref.~\onlinecite{Baran2024Feb}). Furthermore, due to the two-orbital structure of the SI there is a left-over phase from the four tunneling amplitudes that cannot be gauged away and which may impact negatively on the robustness of the spin-triplet. Altogether, this makes the basic DSD units vulnerable to mesoscopic fluctuations, which seem hard to circumvent with gate defined quantum dots and islands. At this point, we can only speculate that fluctuations could be reduced by moving to a super-semi hybrid platform based on highly regular epitaxially grown tunnel barriers~\cite{Junger2019Jul, Thomas2020Jan, Ungerer2024Feb}. Henceforth, we will assume the best-case scenario of symmetric couplings within the DSD unit and focus on the physics that would emerge from assembling many such units into longer chains.

%Experimentally this is not \vb{unreasonable} as the spatial extension of proximitized semiconductor levels can reach several tens or hundreds of nanometers~\cite{Scherubl2020Apr}. 

%In Ref.~\onlinecite{Baran2024Feb}, we pointed out that the spin-singlet-triplet gap may be enhanced by moderately large values of the SI's charging energy and QD-SI tunneling rates. 

We show in Fig.~\ref{fig_2} the behavior of the DSD unit's spin-singlet-triplet gap with increasing QD-SI tunneling rate $\Gamma$, which shows a robust maximum corresponding to the crossing between the two lowest-lying spin-singlet states. \vb{This maximum is located at  $\Gamma\simeq 3.5\Delta$ and reaches around $0.6\Delta$ for $\tilde{L}=2$ (converging to a slightly larger value for $\tilde{L}\geq 3$ surrogates)}. At zero coupling, the difference between the lowest-lying spin-singlets is the presence of a broken Cooper pair with an energy cost of 2$E_{\text{qp}}$. This becomes favored by a strong enough tunneling rate that  encourages the states with single QD occupation to hybridize with the SC quasiparticle excitations (and also allows for empty/doubly occupied QDs).

Furthermore, Fig.~\ref{fig_2} indicates that in the hybridization regime where the DSD's spin-triplet character is the most robust, the QD and SI components contribute democratically to the total spin $S=1$. The physical picture is that of a highly correlated DSD unit, fundamentally different from the weak \vb{QD-SI} coupling scenario where each QD carries a \vb{well defined} spin-1/2 moment. \vb{For weak QD-SI couplings, the DSD chain would thus map well to the alternating ferromagnetic-antiferromagnetic Heisenberg chain of spins-1/2. With increasing antiferromagnetic coupling (the analogue of the inter-dot coupling $t_d$), this model is known to experience a continuous crossover between the Haldane phase and a dimerized phase  \cite{Hida1992Feb,Hida1993May}. }

\begin{figure}[ht!]
\includegraphics[width=\columnwidth]{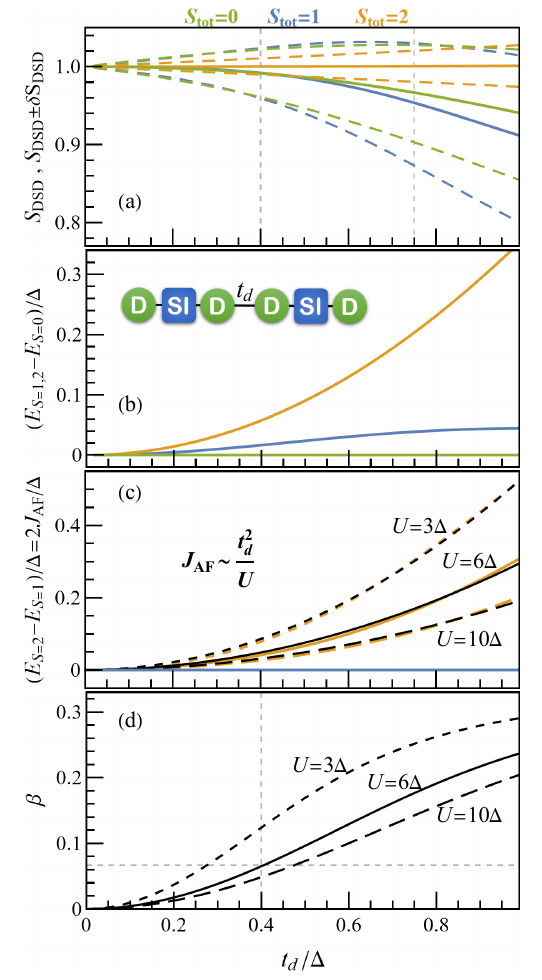}
\caption{(a) Effective total spin $S_{\text{DSD}} $ (continuous lines), defined for each DSD unit by $S_{\text{DSD}}(S_{\text{DSD}}+1)=\langle \vec{S}^{\, 2}_{\text{DSD}}\rangle$, and its estimated range of fluctuations (dashed lines) versus the inter-dot coupling $t_d$, for the $N_{\text{DSD}}=2$ dimer in its lowest-lying  $S_{\text{tot}}=0,1,2$ states.  (b) Spin-triplet and spin-quintuplet excitation energies versus the inter-dot coupling $t_d$. (c-d) BLBQ parameters $J_{\text{AF}}=(E_{S=2}-E_{S=1})/2$ and $\beta$ resulting from fitting the dimer's energy spectrum (black curves). \vb{For the fitting of $J_{\text{AF}}$, see also \cite{blbq_fit}. The values referred in the main text are indicated by dashed grid lines.} The other parameters are $U=6\Delta$ (in a,b), $\epsilon_d=-U/2$, $\Gamma=3\Delta$, $E_c=2\Delta$. For each SI we employed the $\tilde{L}=2$ surrogate.}
\label{fig_3}
\end{figure}

\subsection{The DSD-DSD dimer}

Moving on to the simplest DSD chain, i.e. the $N=2$ dimer, \new{we focus on its lowest lying total-spin-singlet, triplet and quintuplet states, indicated by different colors in Fig.~\ref{fig_3}a-c. For all cases,} we confirm in Fig.~\ref{fig_3}a the gradual breakdown of the spin-triplet character of each DSD unit with increasing inter-dot coupling $t_d$. Here, we show each unit's effective total spin $S_{\text{DSD}}$ defined by $S_{\text{DSD}}(S_{\text{DSD}}+1)=\langle \vec{S}^{\, 2}_{\text{DSD}}\rangle$, together with its range of fluctuations $S_{\text{DSD}}\pm\delta S_{\text{DSD}}$ that reproduces the spread of the total spin squared between $\langle \vec{S}^{\, 2}_{\text{DSD}}\rangle \pm \sqrt{\langle (\vec{S}_{\text{DSD}}^{\,2})^2\rangle-\langle \vec{S}_{\text{DSD}}^{\,2}\rangle^2 }/2$. The total spin of each DSD unit is understood to be $\vec{S}_{\text{DSD}}=\vec{S}_{\text{QD,L}}+\vec{S}_{\text{SI}}+\vec{S}_{\text{QD,R}}$.

The behavior of the DSD-dimer's low-lying energy spectrum interpolates between the AFH-specific scaling $ E_{S=1}-E_{S=0}=(E_{S=2}-E_{S=1})/2=J_\text{AF}$ at low $t_d$, and the AKLT-like scenario with degenerate spin-singlet and triplet ground states at large $t_d$ (see also the discussion in Sec.~\ref{idealchains} above). For intermediate values of $t_d$ we use the BLBQ prescription $E_{S=1}-E_{S=0}=J_\text{AF}(1-3\beta)$, $E_{S=2}-E_{S=1}=2J_\text{AF}$ to obtain the $J_\text{AF}$ and $\beta$ parameters of the effective BLBQ model, see Eq.~(\ref{blbq}). The fitting results are shown in Fig.~\ref{fig_3}c,d, with the effective antiferromagnetic coupling $J_\text{AF}$ following the superexchange scaling $J_{\text{AF}}\sim t_d^2/U$ to a good extent \cite{blbq_fit}, and with the biquadratic term $\beta$ interpolating smoothly between $\beta_{\text{AFH}}=0$ (at small $t_d$) and $\beta_{\text{AKLT}}=1/3$ (at large $t_d$).

\subsection{Longer DSD chains}

\subsubsection{Energy spectra}

\begin{figure}[ht!]
\includegraphics[width=\columnwidth]{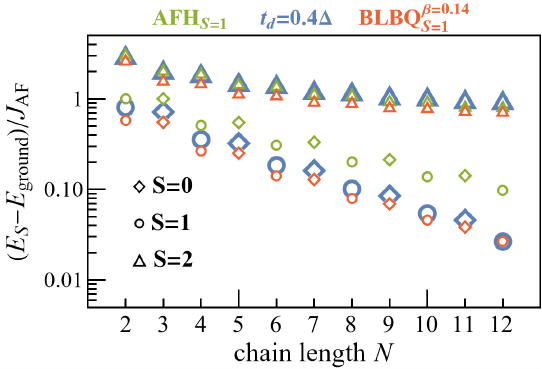}
\caption{Spin energy gaps for the DSD, AFH and BLBQ($\beta=0.14)$ chains relative to the spin-singlet (triplet) ground state for even (odd) chain lengths $N$. The DSD chain parameters are $U=-2\epsilon_d=6\Delta$, $\Gamma=3\Delta$, $E_c=2\Delta$, $t_d=0.4\Delta$ and $J_\text{AF}=0.0205\Delta$. For each SI we employed the $\tilde{L}=2$ surrogate.}
\label{fig_4}
\end{figure}

When increasing to $N\geq 3$ two main signatures of the Haldane phase become manifest, namely the exponential decay of the spin-triplet-singlet gap (with alternating singlet and triplet ground states for even and odd $N$) and the convergence of the spin-quintuplet excitation energy to the corresponding Haldane gap, see Fig.~\ref{fig_4}. Note that a faster breakdown of the spin-triplet character is to be expected for the DSD units in the bulk of $N\geq 3$ chains, as each unit now interacts with both its left and right neighbors. This is apparent in Fig.~\ref{fig_4}, where a value of $\beta_{N=12}=0.14$ within the BLBQ model is needed to reproduce well the DSD spin-gaps for $N=12$, about twice as large when compared to $\beta_{N=2}\simeq 0.07$ required for the $N=2$ DSD-dimer at the same $t_d=0.4\Delta$, cf. also Fig.~\ref{fig_3}d.

\subsubsection{Spin densities and edge states}

\begin{figure}[ht!]
\includegraphics[width=\columnwidth]{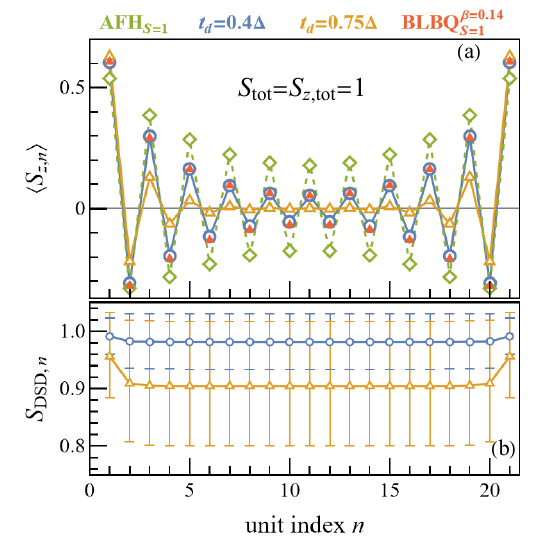}
    \caption{(a) Average spin $\langle S_z\rangle$ along the $N=21$ DSD, AFH and BLBQ($\beta=0.14)$ chains in the $S_{\text{tot}}=S_{z,\text{tot}}=1$ ground state. (b) Effective total spin $S_{\text{DSD},n}$ defined for each DSD unit $n=1,...,21$ by $S_{\text{DSD}}(S_{\text{DSD}}+1)=\langle \vec{S}^{\, 2}_{\text{DSD}}\rangle$, together with its estimated fluctuation range. The other parameters are $U=-2\epsilon_d=6\Delta$, $\Gamma=3\Delta$, $E_c=2\Delta$. For the SI we employed the $\tilde{L}=2$ surrogate.}
\label{fig_5}
\end{figure}

\begin{figure}[ht!]
\includegraphics[width=\columnwidth]{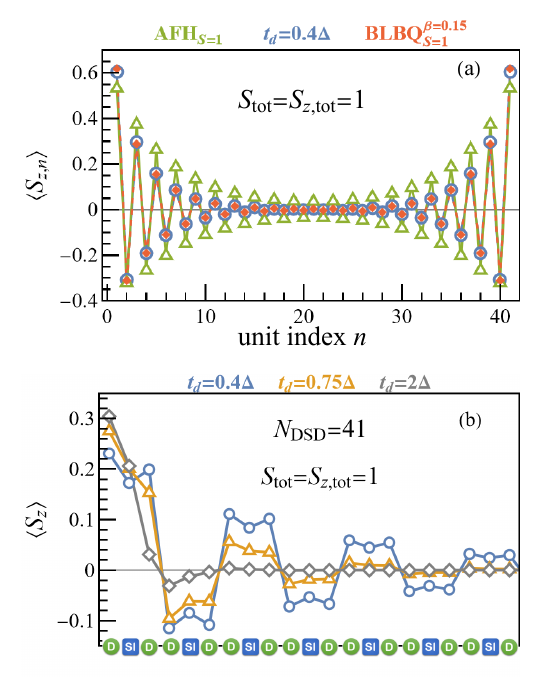}
    \caption{(a) Average spin $\langle S_z\rangle$ along the $N=41$ DSD, AFH and BLBQ($\beta=0.15)$ chains in the $S_{\text{tot}}=S_{z,\text{tot}}=1$ ground state. (b) Detailed average spin $\langle S_z\rangle$ for each QD and SI in the first seven units of the $N=41$ DSD chain, in its $S_{\text{tot}}=S_{z,\text{tot}}=1$ ground state and for various inter-dot couplings $t_d$. The other parameters are $U=-2\epsilon_d=6\Delta$, $\Gamma=3\Delta$, $E_c=2\Delta$. For each SI we employed the $\tilde{L}=2$ surrogate.}
\label{fig_6}
\end{figure}

Even longer chains show clear  signatures specific to the spin-1/2 edge fractionalization \cite{White1993Aug}, see Fig.~\ref{fig_5}a. Here, the average spin in the $S_{\text{tot}}=S_{z,\text{tot}}=1$ ground state of the $N=21$ DSD chain displays the characteristic staggered profile decaying in amplitude away from the edges. This decay is correlated with the strength of the double-dot spin-singlet bonds, being weakest at small $t_d$, i.e. in the AFH regime, and strongest at large $t_d$, i.e. in the AKLT-like \vb{dimerized} regime, where only the end-spins-1/2 survive and there is no bulk magnetization. When going towards the latter regime by progressively increasing $t_d$, the double-dots tighten up into spin-singlets and the DSD units across the entire chain experience increasingly stronger fluctuations, gradually losing their spin-1 character, cf. Fig.~\ref{fig_5}b and see also the discussion around 
Fig.~\ref{fig_3}a.

The increasing-$t_d$ effects on the edge states are shown in Fig.~\ref{fig_6} for a larger $N=41$ DSD chain. In this figure only, we plot the detailed spin distribution for each individual QD and SI instead of that corresponding to entire DSD units. For small to moderate $t_d$, i.e. close to the AFH regime, each QD and SI are seen to contribute a similar amount to the average spin projection $\langle S_z\rangle$ of a DSD unit. This is in agreement with the expectation from the previous analysis of a single DSD unit, see the discussion around Fig.~\ref{fig_2}. At large $t_d$, i.e. in the AKLT-like \vb{dimerized} regime, the bulk average spin-density becomes strongly suppressed as all double-dots are tightly bound into spin-singlet \vb{dimers}. What remains is an effective spin-1/2 moment localized on the outermost QD-SI block, with the individual QD and SI contributions close to their values in an isolated QD-SI system (around 0.3 and, respectively, 0.2 for the chosen parameters).

\subsubsection{String order parameters}

\begin{figure}[ht!]
\includegraphics[width=\columnwidth]{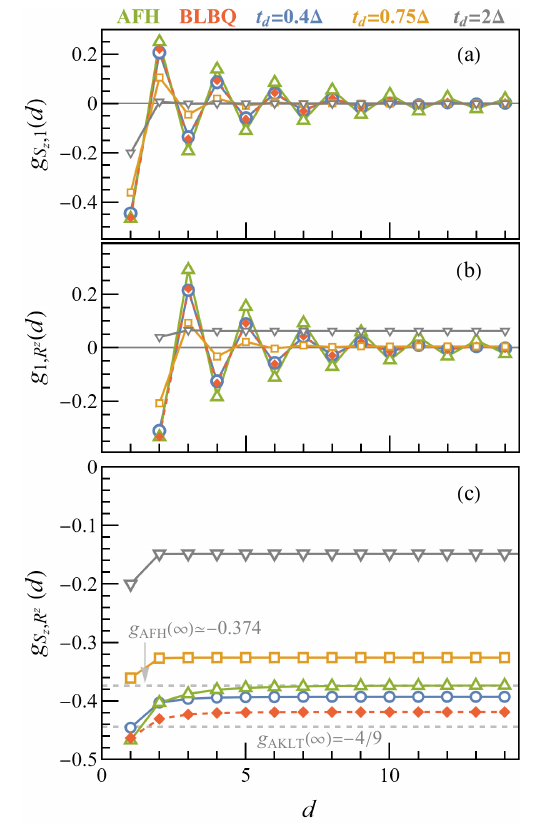}
\caption{(a) Average spin $\langle S_z\rangle$ along the $N=41$ DSD, AFH and BLBQ($\beta=0.15)$ chains in the $S_{\text{tot}}=S_{z,\text{tot}}=1$ ground state. (b-d) Correlation functions $g_{\mathcal{O},\mathcal{U}}(d)$ of Eq.~(\ref{correlator}) for $\mathcal{O}=S_z$ and $\mathcal{U}=1$ (b), $\mathcal{O}=1$ and $\mathcal{U}=R^z$ (c), $\mathcal{O}=S_z$ and $\mathcal{U}=R^z$ (d), in the bulk (middle third, $p=14$) of the $N=41$ chain. The other parameters are $U=-2\epsilon_d=6\Delta$, $\Gamma=3\Delta$, $E_c=2\Delta$. For the SI we employed the $\tilde{L}=2$ surrogate.}
\label{fig_7}
\end{figure}

The relatively long $N=41$ chain features a bulk region large enough to accommodate a sound investigation of various  correlation functions relevant to the Haldane phase, cf. Fig.~\ref{fig_7}.  The correlators adequate here are string order parameters \cite{Pollmann2012Feb} of the form
\begin{equation}\label{correlator}
g_{\mathcal{O},\, \mathcal{U}}(d)=\left\langle\hat{\mathcal{O}}_p\left(\prod_{j=p+1}^{p+d-1}\hat{\mathcal{U}}_j\right)\hat{\mathcal{O}}_{p+d}\right\rangle~,
\end{equation}
which probe the transformation behaviour of the bulk under a symmetry $\mathcal{U}$, e.g. a spin rotation around the $z$ axis with $\pi$, $R^z=\exp(i\pi S_z)$.

For $\mathcal{O}=S_z$ and $\mathcal{U}=1$, Eq.~(\ref{correlator}) reduces to the spin-spin correlation function $g_{S_z,1}=\langle \hat{S}_{z,p}\, \hat{S}_{z,p+d}\rangle$ which is expected to be short ranged as there is no spontaneous breaking of the rotational symmetry in the Haldane phase \cite{Haldane1983Feb}. This behavior is confirmed in Fig.~\ref{fig_7}a.

For $\mathcal{O}=1$ and $\mathcal{U}=R^z$ one deals with the pure-string correlator $g_{1,R^z}$ which is non-zero at large $d$ in the case of topologically trivial configurations \cite{Pollmann2012Feb}. In Fig.~\ref{fig_7}b, as long as $t_d$ is not too large we find the pure-string correlator $g_{1,R^z}$ to decay as expected for a Haldane phase where the bulk $SO(3)$ symmetry fractionalizes into the  $SU(2)$ edge-symmetry. For a larger value like $t_d=2\Delta$ where the DSD spin-1 character is mostly washed out, this pure-string correlator acquires a \new{visibly finite} positive value, signaling a trivial configuration. 

The Haldane phase features a hidden antiferromagnetic order \cite{Kennedy1992Jan} that may be revealed by employing the non-local  string order parameter $g_{S_z,R^z}$ obtained from Eq.~(\ref{correlator}) upon setting $\mathcal{O}=S_z$ and $\mathcal{U}=R^z$ \cite{DenNijs1989Sep,Pollmann2012Feb}. This may be viewed as a standard two-point spin-spin correlator that only picks up a $\pm$ sign from the (non-locally) alternating $S_z=\pm1$ spins while ignoring all $S_z=0$ contributions in between. Fig.~\ref{fig_7}c shows that the DSD chain features a well-defined string order parameter, with a value close to its AFH and BLBQ counterparts for the moderate $t_d=0.4\Delta$. With increasing $t_d$, we notice how the DSD string correlator $|g_{S_z,R^z}(\infty)|$ begins to decrease, e.g. reaching a significantly reduced value of 0.15 at $t_d=2\Delta$ for which the above mentioned pure-string correlator $g_{1,R^z}$ \vb{also had long-range order}.

\vb{When interpreting the order parameters' behavior, it is important to realize that an increasing inter-dot coupling $t_d$ leads to the build-up of density fluctuations which gradually erode the spin-1 character of the DSD units. 
%However, the Haldane phase can be unstable to such density fluctuations \cite{Anfuso2007Aug,Moudgalya2015Apr,Verresen2021Feb}. 
\new{Strictly speaking,} the original SO(3) symmetry of an \new{isolated} spin-1 DSD unit is extended to SU(2) at \new{any finite} $t_d$. \new{At large $t_d$}, the above string orders lose their distinguishing power, with $g_{S_z,R^z}$ and $g_{1,R^z}$ both acquiring a long-range order. Although the Haldane phase
is adiabatically connected to a trivial state, its characteristic phenomena remain parametrically stable (i.e. over a large part of parameter space even when the bulk is in a trivial phase~\cite{Anfuso2007Aug,Moudgalya2015Apr,Verresen2021Feb}).}

In the remainder of this work, we will limit ourselves to the moderate value $t_d=0.4\Delta$ which \new{shows clear signatures of the} Haldane phase.

\subsubsection{Entanglement entropy}

\begin{figure}[ht!]
\includegraphics[width=\columnwidth]{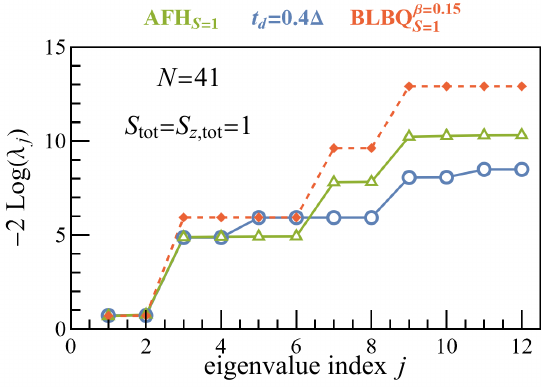}
\caption{ Entanglement spectrum for a bipartition into $(N_\text{L},N_{\text{R}})=(20,21)$ of the  $N=41$ DSD, AFH and BLBQ chains in the $S_{\text{tot}}=S_{z,\text{tot}}=1$ ground state. The DSD parameters are $U=-2\epsilon_d=6\Delta$, $\Gamma=3\Delta$, $E_c=2\Delta$, $t_d=0.4\Delta$ (a) and $t_d=0.75\Delta$(b). For the SI we employed the $\tilde{L}=2$ surrogate.}
\label{fig_8}
\end{figure}

Even in the absence of string order parameters, topological phases can be characterized by their ``entanglement spectrum", obtained upon performing a bipartite cut of the system, tracing out one part and diagonalizing the reduced density matrix of the other \cite{Li2008Jul,Levin2006Mar,Kitaev2006Mar,Turner2010Dec}. \vb{Below, we denote by $\lambda_j$ the Schmidt eigenvalues that square to the eigenvalues of the reduced density matrix.} In particular, the Haldane phases of integer spin chains are characterized by an even degeneracy of the entire entanglement spectrum, caused by the same symmetries protecting the stability of the Haldane phase when applied to the eigenstates of the reduced density matrix \cite{Pollmann2010Feb}. The computation of the entanglement spectrum is straightforward in our MPS approach, and leads to the results shown in Fig.~\ref{fig_8}. All chains under investigation (DSD, AFH and BLBQ) consistently display the even degeneracy required by the Haldane phase (up to minute finite-size effects), and perfectly agree on the dominant pair of eigenvalues. While the higher-lying portions of the AFH and BLBQ's spectra naturally agree on the degeneracy patterns, some deviations occur for the DSD chain due to its underlying microscopic structure. This is to be expected as the entanglement spectrum is a highly sensitive measure of a state's correlations. The corresponding entanglement entropies, computed as $\mathcal{S}=-\sum_j \lambda_j^2\, \text{Log}\, \lambda_j^2$, are $\mathcal{S}_{\text{DSD}}=0.857$, $\mathcal{S}_{\text{AFH}}=0.855$, $\mathcal{S}_{\text{BLBQ}}=0.760$ for Fig.~\ref{fig_8}a.

\subsubsection{Excited states}

\begin{figure}[ht!]
\includegraphics[width=\columnwidth]{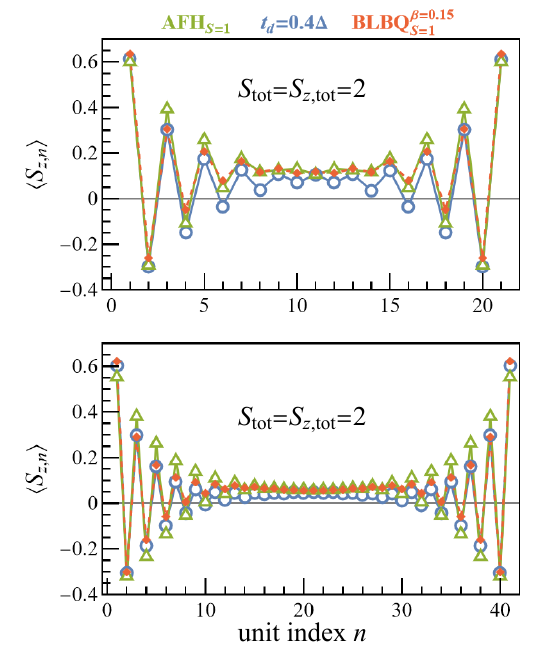}
\caption{ Same as in Figs.~\ref{fig_5}a and \ref{fig_7}a, but for the lowest $S_{\text{tot}}=S_{z,\text{tot}}=2$ excited state.}
\label{fig_9}
\end{figure}

We end this section by examining how well does the mapping of the DSD chain onto the ideal AFH and BLBQ models extend beyond the ground state manifold. In Fig.~\ref{fig_9}, we show the magnetization profile of the lowest-lying $S=2$ state in our longest $N=21$ and $N=41$ chains. All models agree well on the familiar staggered profile giving rise to the two edge-spins-1/2, and also on the bulk acquiring a quasi-uniform spin density (responsible for the spin-1 magnon excitation). The only noticeable quantitative discrepancy is related to the DSD staggered profile extending slightly more into the bulk than in the case of the AFH or BLBQ. While higher-order corrections such
as bicubic exchange couplings could be considered towards reaching a better quantitative agreement of an ideal $S=1$ chain with the DSD chain, this is well beyond the scope of this work.

\section{Conclusions}
\label{sec:conclusions}

The main purpose of this work was to show that the Haldane phase may be realized in a superconductor-semiconductor hybrid platform, more precisely in a chain of repeating QD-SI-QD (DSD) blocks, each exhibiting a robust spin-1 character over a sizeable parameter regime. As long as the coupling between neighboring DSD units was not too strong to destroy their spin-1 character, the basic physics of the DSD chain could be quite well fitted by the bilinear-biquadratic spin-1 Hamiltonian of Eq.(\ref{blbq}) with a biquadratic coefficient $0<\beta<1/3$. In this regime, the DSD chain was found to \new{exhibit clear signatures of the} Haldane phase, such as the presence of characteristic spin-density profiles with effective spins-1/2 at the edges, \new{the long-range order of specific} string correlation functions and the double-degeneracy of the entanglement spectrum. \ff{Our model could be extended, for example, by including an external magnetic field in order to define a singlet-triplet qubit protected from decohence by a Haldane gap~\cite{Jaworowski2017Jul}. In this regime, our DSD unit would be closely related to the setup used in Ref.~\onlinecite{Tsintzis2022Nov} to create poor man's Majorana states; it would thus be interesting to explore a possible cross-over between the Haldane and Majorana physics in these systems.}

\begin{figure}[ht!]
\includegraphics[width=0.66 \columnwidth]{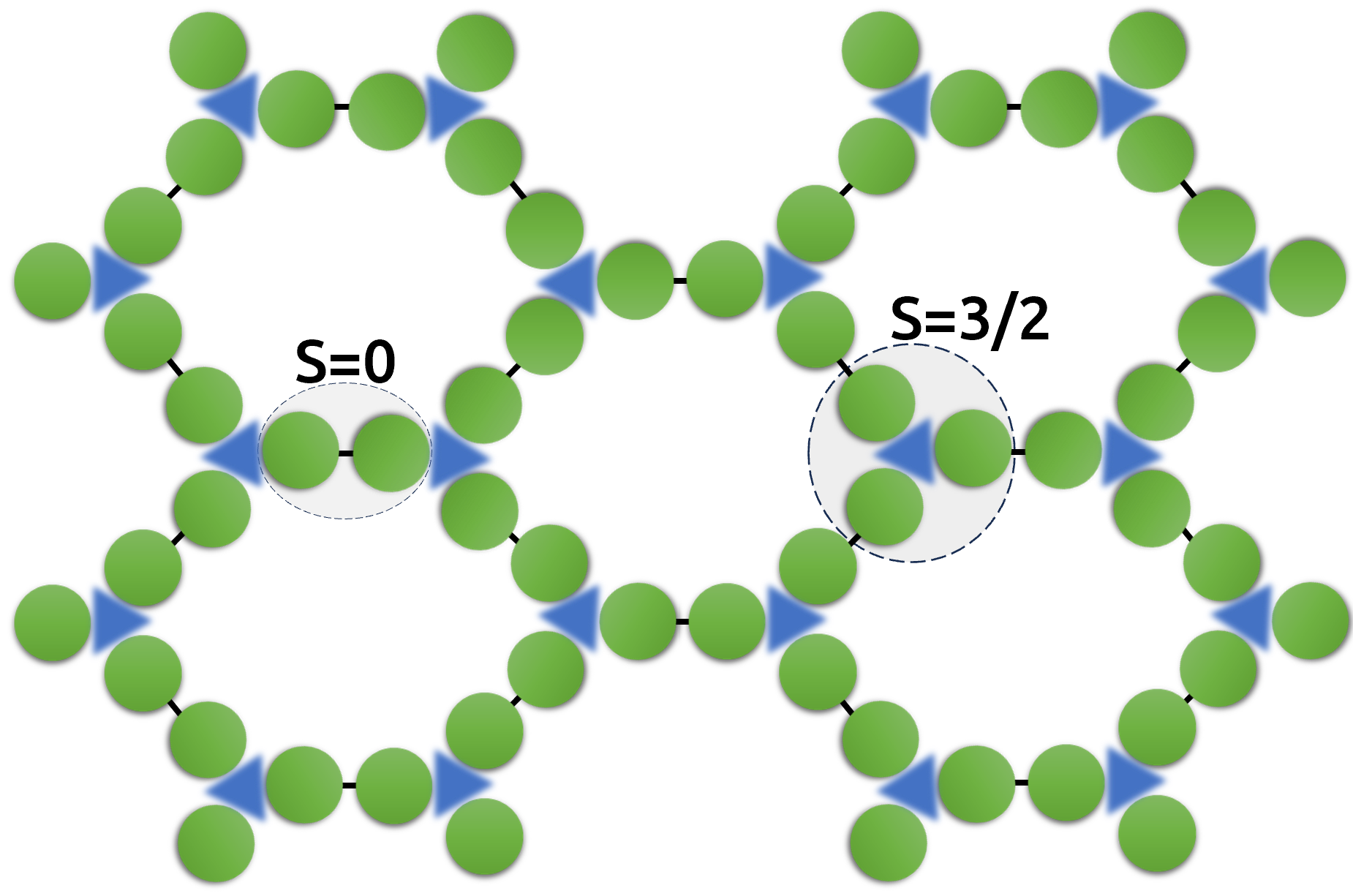}
\caption{Illustrations of the spin-3/2 QD$^3$-SI honeycomb lattice. The QDs (SIs) are indicated green dots (blue triangles).}
\label{fig_10}
\end{figure}

One of the main advantages of the present super-semi platform lies in the ease of designing higher-spin units: by individually tunnel-coupling a number $\mathcal{N}$ of QDs to the same SI we would obtain a robust spin $S=\mathcal{N}/2$ unit \cite{Baran2024Feb}. This could enable the realization of more general spin models in various geometries \cite{Ren2023Dec}. In particular, the generalization of the AKLT state to spins-3/2 on a hexagonal lattice has notable implications for quantum computation \cite{Wei2011Feb,Wei2012Sep}. However, the Heisenberg model on this lattice exhibits N\'eel order and is not in the same phase as the AKLT model \cite{Huang2016Oct}, but a more general bilinear-biquadratic-bicubic model may actually be tuned to an AKLT phase \cite{Huang2013Nov}.  In this context, it would be worthwhile to investigate how the present work generalizes to the analogous QD$^3$-SI hexagonal network depicted in Fig.~\ref{fig_10}. 

\begin{acknowledgments}

\ff{We thank M. Burrello, K. Flensberg and R. S. Souto for stimulating discussions and suggestions}. This work was supported by a grant of the Romanian Ministry of Education and Research, Project No. 760122/31.07.2023 within PNRR-III-C9-2022-I9.
\end{acknowledgments}

\appendix*
\section{QD-SI-QD perturbation theory}
\label{perturbation_theory}
We provide here additional insights from perturbation theory regarding the spin-singlet-triplet competition in the QD-SI-QD system. For a complementary picture, see also Appendix D of Ref.~\onlinecite{Bacsi2023Sep}. Our discussion below will parallel that of Sec. III A in Ref.~\onlinecite{Probst2016Oct}.

For simplicity, we work here in the equivalent BCS picture obtained after transferring the charging term from the SI to the QDs, see Ref.~\onlinecite{Baran2024Feb} for details. As in the main text, we assume that the superconductor is described by an $\tilde{L}=2$ surrogate with both levels coupled to each QD by the same tunneling amplitude $t$. We denote the quasiparticle energy by $E_{\text{qp}}=\sqrt{\Delta
^2+\xi^2}$, with $\xi=\xi_1=-\xi_2>0$ indicating the levels' positions. For convenience we employ the BCS coherence factors $u,v$ satisfying $uv=\Delta/2E_\text{qp}$, $u^2-v^2=\xi/E_{\text{qp}}>0$.

Straightforward 4th order non-degenerate perturbation theory in the tunnel coupling $t$ (implemented for each total spin subspace using the Sneg software \cite{Zitko2011Oct,sneg2}) leads to the spin-singlet-triplet gap
\begin{equation}
\label{deltaTS}
\begin{aligned}
   \delta_{\text{S-T}}&\equiv  \frac{E_{S=1}-E_{S=0}}{t^4}\\&=\frac{16}{\left(U/2+E_{\text{qp}}+E_c\right)^2\left(U+ 2E_{\text{qp}}+4E_c\right) }\\
    &+\frac{64 u^2v^2}{\left(U/2+E_{\text{qp}}+E_c\right)^2\left(U+ 4E_{c}\right) }\\
   &+\frac{64u^2v^2}{\left(U/2+E_{\text{qp}}+E_c\right)^2\left(U+2 E_{\text{qp}}\right) }\\
   &-\frac{16 \left(u^2-v^2\right)^2}{\left(U/2+E_{\text{qp}}+E_c\right)^2\,2E_{\text{qp}}}~. 
   \end{aligned}
\end{equation}

The perturbative expansion may be visualized in terms of spin-exchange processes with the matrix
element of each process being
weighted by the inverse product of the virtual excitation
energies. A final state with exchanged spins may be reached via
intermediate virtual states connected by four
tunnelling events between the QDs and the superconductor.

In processes that involve only virtual excitations
on the QDs the two initial electrons have to be
swapped, leading to an overall sign that energetically favors the spin-singlet state. This is the case for the first three terms in Eq.~(\ref{deltaTS}). It is however possible to also exchange the spins without
anticommutation signs through processes in which a hole is involved. This kind of processes will energetically favor the spin-triplet state, leading to the appearance of the last term in Eq.~(\ref{deltaTS}).

The ratio between
triplet-favoring and singlet-favoring processes is given schematically by $1+E_{\text{QDs}}/E_{\text{qp}}$, where $E_{\text{QDs}}\sim U+E_c$ is a typical excitation energy in the QD subsystem. When it becomes advantageous to perform the spin-exchange by virtually exciting the superconductor instead of the QDs, a spin-triplet ground state may emerge.

Notice however that there is a certain amount of destructive interference in  last term of Eq.~(\ref{deltaTS}). Namely, its subset of processes involving all possible excitations on both superconducting levels (of the type $u_1v_1u_2v_2=u^2v^2$) will favor the spin-singlet instead. This effect completely suppresses the last term of Eq.~(\ref{deltaTS}) in the zero-bandwidth limit of degenerate levels $\xi=0$: for the spin-triplet to be the ground-state,
at least two distinct levels with enough separation are required in the superconductor.

\begin{figure}[ht!]
\includegraphics[width=\columnwidth]{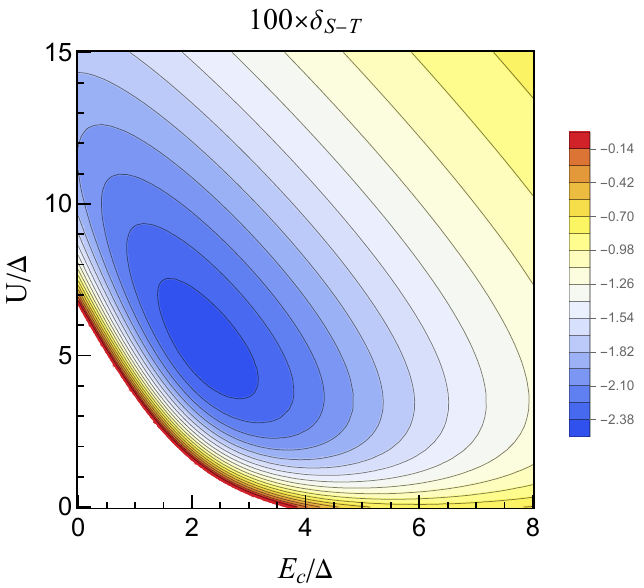}
\caption{Spin-singlet-triplet gap of Eq.~(\ref{deltaTS}) versus $U$ and $E_c$. The other parameters are $\Delta=1$, $\epsilon_d=-U/2$, $\xi=1.31\Delta$. Only the region with a spin-triplet ground state is colored.}
\label{fig_11}
\end{figure}

Finally, we consider the large-$U$ limit 
\begin{equation}
  \delta_{\text{S-T}}=  -\frac{32\, \xi^2}{U^2 \,E_{\text{qp}}^3}+\frac{64 \left(3+{2\, E_c\, \xi^2}/E_{\text{qp}}^3\right)}{U^3}+\mathcal{O}\left(U^{-4}\right)
\end{equation}
and the large-$E_c$ limit 
\begin{equation}
    \delta_{\text{S-T}}= \frac{8}{E_c^2}\left[-\frac{\xi^2}{ \,E_{\text{qp}}^3}+\frac{2\,\Delta^2
    }{E_{\text{qp}}^2(U+2E_{\text{qp}})}\right]+\mathcal{O}\left(E_c^{-3}\right)
\end{equation}
which are both in agreement with the above considerations (and with the conclusions of Ref.~\onlinecite{Bacsi2023Sep}) regarding the existence of a finite bandwidth threshold for establishing the spin-triplet ground state. Note however that the numerical examples considered in the main text (with $U=6\Delta, E_c=2\Delta$) do not fall under any of these limits, but are chosen instead to ensure the maximum spin-singlet-triplet gap, see Fig.~
\ref{fig_11}.

\bibliography{apssamp}

\begin{thebibliography}{82}
\bibitem [{blb()}]{blbq_fit}%
  \BibitemOpen
  \href@noop {} {\bibinfo {title} {{More precisely, we fitted the $t_d^2$
  dependency of the DSD-dimer's spin-quintuplet-triplet gap for three different
  values of $U$, obtaining $J_{\text{AF}}=\gamma t_d^2/U$ with the
  proportionality factor $\gamma$ only showing a weak $U$ dependence,
  $\gamma=0.8,0.9, 0.97$ for $U/\Delta=3,6,10$.}}}\BibitemShut {Stop}%
\end{thebibliography}

\end{document}